\documentclass{nature}
\usepackage{graphicx}
\usepackage{makecell}
\usepackage{subcaption}
\makeatletter
\let\saved@includegraphics\includegraphics
\AtBeginDocument{\let\includegraphics\saved@includegraphics}
\renewenvironment*{figure}{\@float{figure}}{\end@float}
\makeatother

\usepackage[x11names]{xcolor}
\usepackage{amssymb}

\newcommand{\mn}{{Mon. Not. R. Astron. Soc.}}
\newcommand{\mnras}{\mn}
\newcommand{\aj}{{Astron. J.}}
\newcommand{\apj}{{Astrophys. J.}}
\newcommand{\apjl}{{Astrophys. J. Lett.}}
\newcommand{\apjs}{{Astrophys. J. Supp.}}

\newcommand{\aap}{{Astron. Astrophys.}}
\newcommand{\nat}{{Nat.}}
\newcommand{\natas}{{Nat. Ast.}}

\newcommand{\pasp}{{Pub. Ast. Soc. Pac.}}

\newcommand{\ssr}{Space Science Reviews}

\title{A rotating white dwarf shows different compositions on its opposite faces}

\author{Ilaria Caiazzo\footnote{email: ilariac@caltech.edu} $^1$, Kevin B. Burdge$^{2,3}$, Pier-Emmanuel Tremblay$^4$, James Fuller$^1$, Lilia Ferrario$^5$, Boris T. G\"{a}nsicke$^4$, J. J. Hermes$^6$, Jeremy Heyl$^7$, Adela Kawka$^8$, S. R. Kulkarni$^1$, Thomas R. Marsh$^4$, Przemek Mr{\'o}z$^9$, Thomas A. Prince$^1$, Harvey B. Richer$^7$, Antonio C. Rodriguez$^1$, Jan van Roestel$^{1,10}$, Zachary P. Vanderbosch$^1$, St\'{e}phane Vennes$^5$, Dayal Wickramasinghe$^5$, Vikram S. Dhillon$^{11,12}$, Stuart P. Littlefair$^{11}$, James Munday$^4$, Ingrid Pelisoli$^4$, Daniel Perley$^{13}$, Eric C. Bellm$^{14}$, Elm\'e Breedt$^{15}$, Alex J. Brown$^{11}$, Richard Dekany$^{16}$, Andrew Drake$^1$, Martin J. Dyer$^{11}$, Matthew J. Graham$^1$, Matthew J. Green$^{17}$, Russ R. Laher$^{18}$, Paul Kerry$^{11}$, Steven G.  Parsons$^{11}$, Reed L. Riddle$^1$, Ben Rusholme$^{18}$, Dave I. Sahman$^{11}$}

\begin{document}

\maketitle

\begin{affiliations}
 \item Division of Physics, Mathematics and Astronomy, California Institute of Technology, Pasadena, CA 91125, USA
 \item Department of Physics, Massachusetts Institute of Technology, Cambridge, MA 02139, USA
 \item Kavli Institute for Astrophysics and Space Research, Massachusetts Institute of Technology, Cambridge, MA 02139, USA
 \item Department of Physics, University of Warwick, Gibbet Hill Road, Coventry CV4 7AL, UK
 \item Mathematical Sciences Institute, The Australian National University, ACT 0200, Australia
 \item Department of Astronomy \& Institute for Astrophysical Research, Boston University, Boston 02215, USA
 \item Department of Physics and Astronomy, University of British Columbia, Vancouver, BC, V6T1Z1, Canada
 \item International Centre for Radio Astronomy Research, Curtin University, GPO Box U1987, Perth, WA 6845, Australia
 \item Astronomical Observatory, University of Warsaw, Al. Ujazdowskie 4, 00-478, Warszawa, Poland
 \item Anton Pannekoek Institute, University of Amsterdam, Postbus 94249, NL-1090 GE Amsterdam, The Netherlands
 \item Department of Physics and Astronomy, University of Sheffield, Sheffield, S3 7RH, UK
 \item Instituto de Astrof\'isica de Canarias, E-38205 La Laguna, Tenerife, Spain
 \item Astrophysics Research Institute, Liverpool John Moores University, IC2, Liverpool Science Park, 146 Brownlow Hill, Liverpool L3 5RF, UK
 \item Department of Astronomy, University of Washington, Seattle, WA 98195, USA
 \item Institute of Astronomy, University of Cambridge, Madingley Road, Cambridge CB3 0HA, UK
 \item Caltech Optical Observatories, California Institute of Technology, Pasadena, CA  91125
 \item Department of Astrophysics, School of Physics and Astronomy, Tel Aviv University, Tel Aviv 6997801, Israel
 \item IPAC, California Institute of Technology, 1200 E. California Blvd, Pasadena, CA 91125, USA

\end{affiliations}

\begin{abstract}
White dwarfs, the extremely dense remnants left behind by most stars after their death, are characterised by a mass comparable to that of the Sun compressed into the size of an Earth-like planet. In the resulting strong gravity, heavy elements sink toward the centre and the upper layer of the atmosphere contains only the lightest element present, usually hydrogen or helium\cite{1958whdw.book.....S,1986ApJS...61..197P}. Several mechanisms compete with gravitational settling to change a white dwarf's surface composition as it cools\cite{1987fbs..conf..319F}, and the fraction of white dwarfs with helium atmospheres is known to increase by a factor $\sim2.5$ below a temperature of about 30,000~K\cite{1986ApJS...61..305G,1986ApJ...309..241L,1991ApJ...371..719M,2006AJ....132..676E,2020ApJ...901...93B}; therefore, some white dwarfs that appear to have hydrogen-dominated atmospheres above 30,000~K are bound to transition to be helium-dominated as they cool below it. 
Here we report observations of ZTF J203349.8+322901.1, a transitioning white dwarf with two faces: one side of its atmosphere is dominated by hydrogen and the other one by helium. This peculiar nature is likely caused by the presence of a small magnetic field, which creates an inhomogeneity in temperature, pressure or mixing strength over the surface\cite{2014Natur.515...88V,tremblay:15,2020MNRAS.492.3540C}. ZTF J203349.8+322901.1 might be the most extreme member of a class of magnetic, transitioning white dwarfs -- together with GD 323\cite{2005ApJ...623.1076P}, a white dwarf that shows similar but much more subtle variations. This new class could help shed light on the physical mechanisms behind white dwarf spectral evolution. 
\end{abstract}

\begin{figure}
    \centering
    \includegraphics[width=\textwidth]{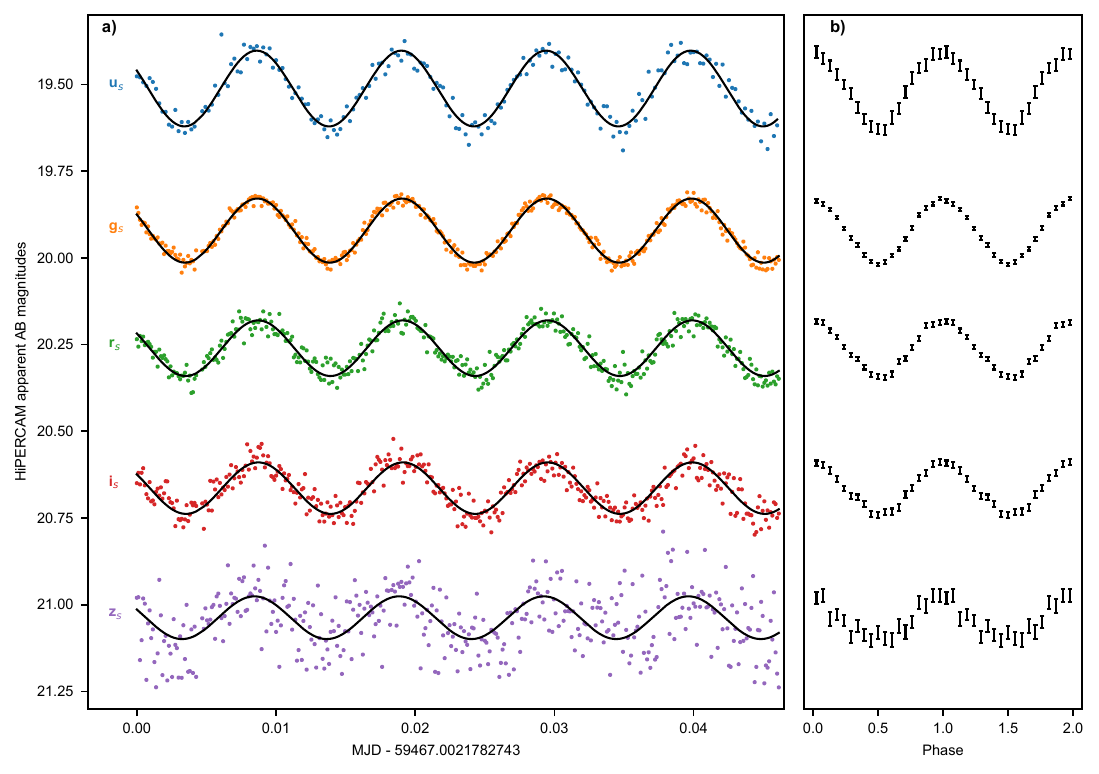}
    \caption{{\bf Janus HiPERCAM lightcurve.} {\bf a)} The five-colour HiPERCAM light curve of Janus with filters, from top to bottom, $u_{\mathrm{s}}$, $g_{\mathrm{s}}$, $r_{\mathrm{s}}$, $i_{\mathrm{s}}$ and $z_{\mathrm{s}}$. The redder filters are progressively fainter because the white dwarf is hot. Also, the amplitude of variation decreases toward the red from 22\% peak-to-peak in $u_{\mathrm{s}}$ to 12\% in $z_{\mathrm{s}}$ (see Table~\ref{tab:fitpar}). The solid black lines show sinusoidal fits to the light curves.  (\textbf{b}) The binned HiPERCAM light curves phase-folded at a period of 14.97 minutes. Phase zero corresponds to the peak of the light curve and the error bars indicate 1$\sigma$ errors.}
    \label{fig:lc}
\end{figure}

\begin{figure}
    \centering
    \includegraphics[width=\textwidth]{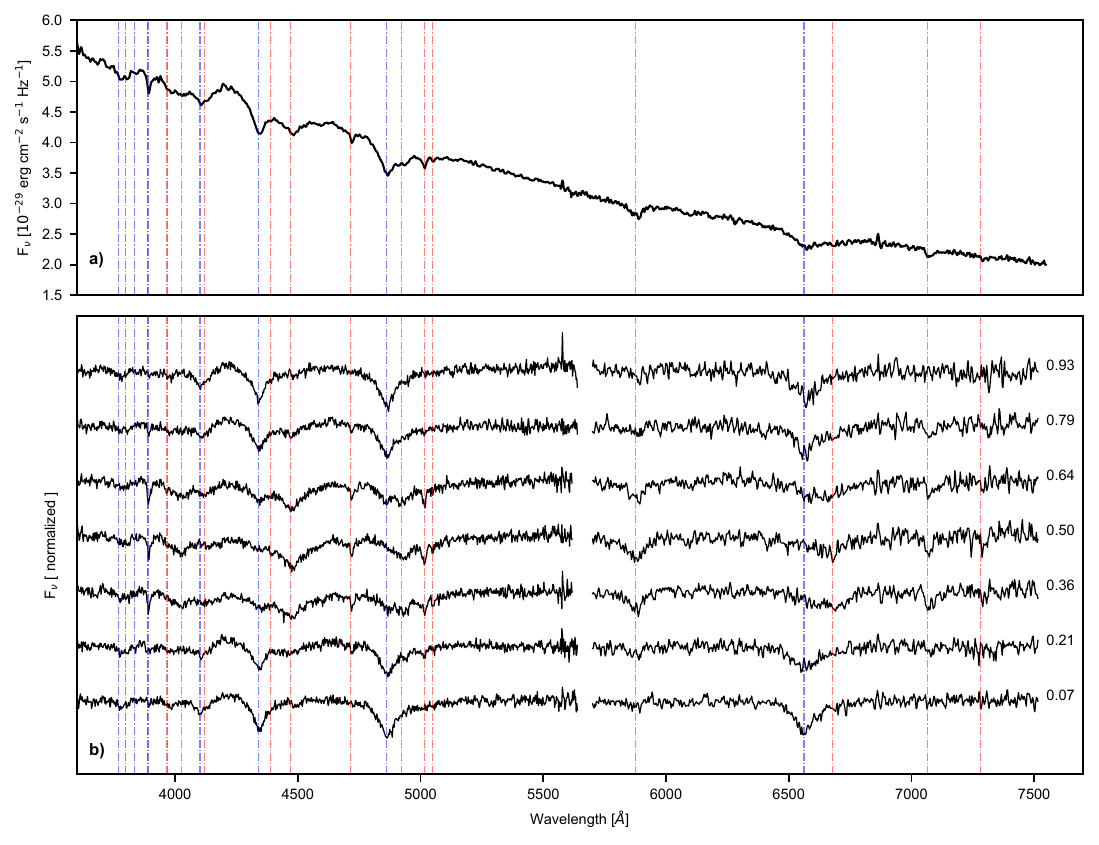}
    \caption{{\bf LRIS spectra.} {\bf a)} Phase-averaged spectrum of Janus; the blue vertical lines highlight the position of the hydrogen Balmer lines, while the red ones indicate the absorption lines of neutral helium. {\bf b)} Phase-resolved and phase-binned spectra, normalised and shifted vertically for clarity. The numbers on the right indicate the centre of each phase bin. Phase 0 corresponds to the maximum in the light curve (Fig.~\ref{fig:lc}). Helium lines are absent near phase 0 while hydrogen lines are absent near phase 0.5, indicating the varying composition across the surface of the star. The transition is best revealed by the helium line at 3888~\AA\ and the hydrogen H$\beta$ line at 4,861~\AA. }
    \label{fig:pa-sp}
\end{figure}

The white dwarf ZTF J203349.8+322901.1, which we nicknamed Janus after the two-faced Roman god of transition, was found during a search for periodically variable white dwarfs with the Zwicky Transient Facility\cite{2019PASP..131a8002B} (ZTF). Observations with the high-speed imaging photometer \emph{CHIMERA}\cite{2016MNRAS.457.3036H} on the Hale telescope and the quintuple-beam imager HiPERCAM\cite{Dhillon2021} on the Gran Telescopio Canarias revealed a large amplitude sinusoidal light curve with a 14.97 minutes period (see Fig.~\ref{fig:lc}, Table~\ref{tab:fitpar} and Extended Data Fig.~\ref{fig:lc_ztf}). Next, we obtained phase-resolved spectroscopy using the Low-Resolution Imaging Spectrometer (\emph{LRIS})\cite{1995PASP..107..375O} on the Keck I Telescope (see Fig.~\ref{fig:pa-sp}). The white dwarf's spectrum transitions from showing only hydrogen lines at phase $\approx0$ (the phase of maximum brightness in the photometric light curve, Fig.~\ref{fig:lc}) to only helium lines at phase $\approx0.5$ (minimum brightness).

These spectroscopic and photometric variations rule out the hypothesis of Janus being a binary system composed of a white dwarf with a hydrogen-dominated spectrum (usually called a DA white dwarf) and one with a helium-dominated one (called DB). In fact, in order to explain Janus' spectroscopic variability, the binary would have to be eclipsing and, at a 15-minute orbital period, radial velocities would be apparent as Doppler shifts of the order of $\sim1,000$~km/s, corresponding to wavelength shifts of 15 to 30~\AA, which are not detected in Janus' spectra. Additionally, the light curve is sinusoidal and does not show eclipses. Short-period sinusoidal modulations are observed in pulsating white dwarfs; however, the simple change in temperature caused by pulsations cannot explain the change in spectral composition observed in Janus. We therefore conclude that the white dwarf is rotating with a period of 14.97 minutes and that the spectral and photometric changes are caused by a variation in composition across the surface. Such short rotation periods are rare for isolated white dwarfs (typical periods range from hours to days\cite{2017ApJS..232...23H}), and they are usually associated with a merger origin because they are often observed in highly magnetised white dwarfs\cite{2021Natur.595...39C,2021ApJ...923L...6K}.

\begin{figure}
    \centering
    \includegraphics[width=0.42\textwidth]{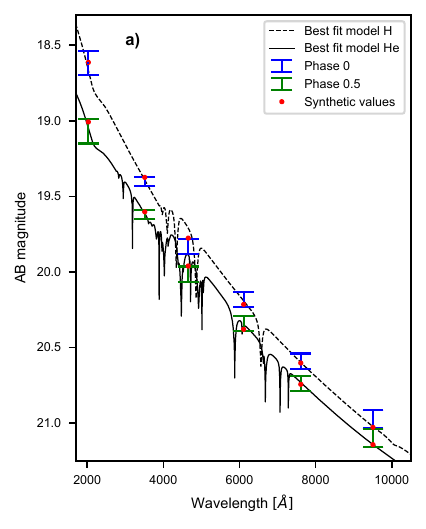}
    \includegraphics[width=0.57\textwidth]{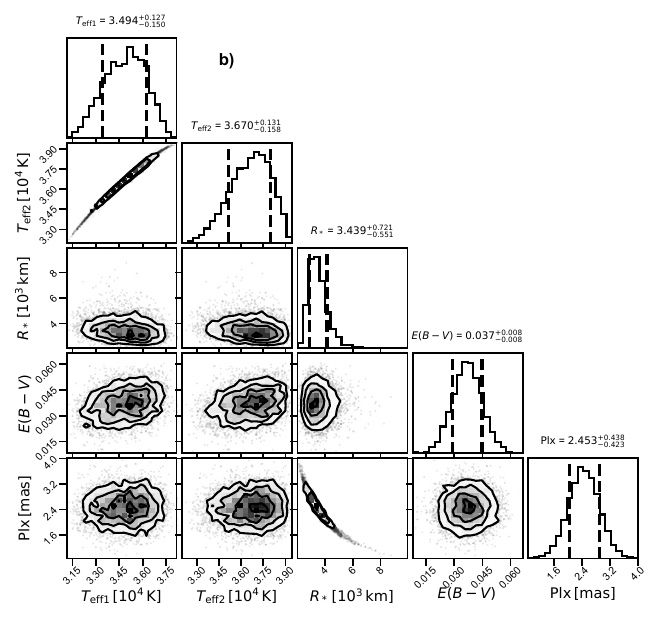}
    \caption{{\bf SED fitting of the two faces.} The plots show the simultaneous fitting of the SED on phase 0 and phase 0.5 with hydrogen- and helium-dominated atmospheres models. {\bf a)}: The blue error bars show the AB magnitude of Janus at the maximum of the light curve (phase 0) in \emph{Swift} $UVW2$ filter and in the HiPERCAM filters, while the green error bars show the minimum (phase 0.5). The black lines show the best-fitting atmosphere models for phase 0 (hydrogen model, dashed line) and for phase 0.5 (helium model, solid line), and the red dots show the corresponding synthetic magnitudes in the different filters. Although the observed magnitudes in both phases are shown with their error bar in absolute calibration, the fitting was performed by fixing the relative amplitudes between phases in each filter with their errors as listed in Table~\ref{tab:fitpar} (see Methods). {\bf b)}: Corner plot for the fitting procedure. We simultaneously fit the SED data on the two faces allowing different temperatures ($T_{\rm{eff}1}$ for phase 0 and $T_{\rm{eff}2}$ for phase 0.5) but assuming the same radius, interstellar extinction and distance (Table~\ref{tab:fitpar}).}
    \label{fig:allfit}
\end{figure}

To constrain the white dwarf's temperature, we obtained photometric measurements in the near UV using the UVOT instrument\cite{2005SSRv..120...95R} on the Neil Gehrels Swift Observatory\cite{2004ApJ...611.1005G}. The peak-to-peak amplitude variation in the UV is more than twice that of the optical, $46\pm8\%$. We compared the observed spectral energy distribution (SED) at maximum and minimum brightness (using the HiPERCAM and \emph{Swift} data) to synthetic atmosphere models (see the Methods section). For phase 0, we used pure-hydrogen atmosphere models, while for phase 0.5 we tried several models with hydrogen to helium mass ratio ranging from $10^{-30}$ to $10^{-3}$ (a higher hydrogen content would result in a detectable H$\alpha$ line in the spectrum), but we found that the composition did not affect the SED fitting at these values. The best fitting models are shown in Fig.~\ref{fig:allfit}, while the properties obtained from the fit are listed in Table ~\ref{tab:fitpar}. We compared the radius and temperature obtained with white dwarf cooling models\cite{2020ApJ...901...93B,2019A&A...625A..87C} to infer the mass of the white dwarf; depending on the core composition, we infer a mass between 1.2 and 1.27 solar masses. While we obtain a temperature of about $\sim35,000$~K for both faces, the absorption lines in the spectra are very weak, as are usually observed in white dwarfs with a much higher surface temperature, close to $50,000$~K.
At such a high temperature, we would expect a steeper SED and a strong He II absorption line at 4,685 \AA, which is absent in Janus' spectrum at phase 0.5. We therefore conclude that the temperature of the white dwarf is close to what we infer from the SED and that the lines are weakened by some other mechanism. As we explain below, the peculiar double-faced nature of the Janus might be due to the presence of a magnetic field on its surface, which could also weaken the absorption lines. In fact, disagreement between temperatures derived from photometry and from spectroscopy is very common in magnetic white dwarfs\cite{2023MNRAS.520.6111H,2015ASPC..493...53R}, although such an effect remains to be demonstrated at the temperature and field strength of Janus.

\begin{figure}
    \centering
    \includegraphics[width=0.7\textwidth]{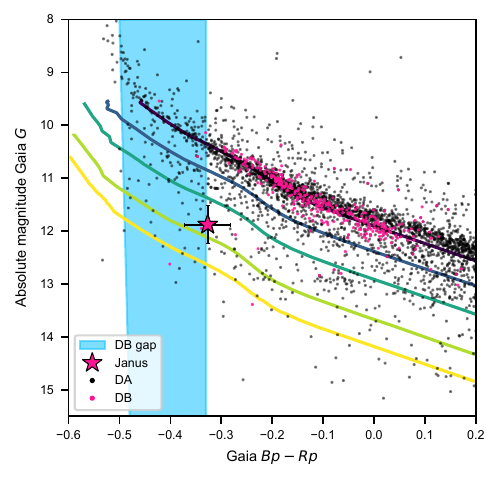}
    \caption{{\bf Gaia colour-magnitude diagram.} Gaia colour-magnitude diagram showing the location of Janus below the main white dwarf track (magenta star). The x-axis depicts the difference between the Gaia Bp and Rp bands, and the y-axis the absolute magnitude in the G filter. White dwarfs that are within 200~pc from Earth and that have a spectral classification in the Montreal Catalogue\cite{2017ASPC..509....3D} are plotted as black dots if DAs and magenta if DBs. Solid lines show theoretical cooling tracks for white dwarfs with masses, from top to bottom, 0.6, 0.8, 1.0, 1.2 and 1.3~M$_\odot$; the atmosphere is assumed to be helium-dominated\cite{2011ApJ...737...28B} and the interior composition to be carbon-oxygen\cite{2020ApJ...901...93B} for $M<1.1$~M$_\odot$ and oxygen-neon\cite{2019A&A...625A..87C} for $M>1.1$~M$_\odot$. The shaded area indicates the range $\approx50,000-30,000$~K according to the same models, which is the location of the DB Gap. Reddening corrections were applied only to Janus. 1$\sigma$ error bars are shown for Janus, and are omitted for the background dots for clarity.}
    \label{fig:cmd}
\end{figure}

The temperature of Janus, $\sim35,000$~K, places it at the red edge of the ``DB gap'', a range of temperatures in which the observed incidence of DB white dwarfs is about 2.5 times smaller compared to DAs than at neighbouring temperatures\cite{1986ApJS...61..305G,1986ApJ...309..241L,2006AJ....132..676E} (see Fig.~\ref{fig:cmd}). Because of the paucity of DBs in the gap, some hot and young white dwarfs with helium-rich atmospheres are bound to transition into DAs as they cool below $\approx50,000$~K. The current explanation for this transition is the upward diffusion of a very thin layer of hydrogen under the influence of gravitational settling.
The reappearance of DBs below the cool end of the gap is interpreted as the consequence of the dilution of the hydrogen layer by the growing convection zone in the underlying helium envelope, provided that the surface hydrogen layer is thin enough (between $10^{-17}$ and $10^{-14}$ solar masses\cite{1987fbs..conf..319F,2016ApJ...833..127M,2020ApJ...901...93B}). The temperature at which this second transition occurs likely depends on the mass of the white dwarf (with higher masses transitioning at higher temperatures) and on the thickness of the hydrogen layer, so instead of a sharp edge, the red end of the DB gap looks like a smooth transition between 35,000 and 25,000~K\cite{2022A&A...658A..79L}. 

Janus could be currently undergoing the second transition at the cool edge of the DB gap. If a small magnetic field is present on the surface of the white dwarf, it would be enough to inhibit convective mixing on part of the white dwarf's surface. In this case, the hydrogen-rich (bright) face of the white dwarf would correspond to the region where the field is strong enough to impede convection, so that the hydrogen layer is preserved. The helium-dominated face, on the other hand, would be where the magnetic field is weaker, and convection strong enough to corrode the thin hydrogen layer. A simple estimate for the minimum magnetic field needed to inhibit convection can be obtained by imposing the magnetic field pressure to equal the gas pressure at an optical depth of about 1; we find that a field of a few tens of kiloGauss (kG) would be enough to stop convection (see Methods). The absorption lines do not show any Zeeman splitting or wavelength shifts, so we know that the magnetic field on either face cannot be higher than a few MG. Alternatively, since the onset of convection due to neutral helium is extremely sensitive to the effective temperature\cite{2019MNRAS.490.1010C}, if the magnetic field induced a small difference in temperature between the two faces, it would be enough to explain the difference in composition: on the hot side, the mixing region is not yet extended into the hydrogen layer and we still see hydrogen, while at the colder side mixing has already diluted the hydrogen and we see only helium. 

Even in the absence of convection, the presence of a small magnetic field on the surface of the star may be able to explain the double-faced nature of Janus. If we assume the magnetic field to be stronger on one side (as could be the case for an offset dipole structure and is witnessed for example in the white dwarf G183-35\cite{2019MNRAS.489.3648K}), the magnetic pressure in the atmosphere would be higher at the pole than on the rest of the surface. As a consequence, the gas pressure and density at the same gravitational potential surface (i.e. at the same radius within the white dwarf) would be lower at the pole, and the ion pressure gradient would cause ions with low mass-to-charge ratio, in this case hydrogen, to diffuse toward the low-pressure region at the pole (see Extended Data Figures~\ref{fig:hd} and \ref{fig:hf}). If the hydrogen content is low enough, below $\sim \! 10^{-14}$ solar masses, the hydrogen accumulated at the poles would cover only part of the stellar surface, forming a hydrogen ocean.

In all these scenarios, the small hydrogen content and surface temperature characterise Janus as a white dwarf currently transitioning out of the DB Gap. As it will cool to lower temperatures, the much stronger convection will succeed in overcoming the magnetic field and destroying the hydrogen layer, transforming the white dwarf into a normal DB.
Spectral variations, although not as extreme as in Janus, have been observed in another white dwarf, GD~323, which is also close to the red edge of the DB Gap\cite{1994ApJ...422..783K} and whose spectrum shows hydrogen and helium absorption features that appear weak for its temperature. GD 323 also exhibits quasi-periodic variations in both hydrogen and helium absorption lines, and the strength of the hydrogen and helium lines is anticorrelated\cite{2005ApJ...623.1076P}. Janus might therefore not be an isolated case but rather the most striking member of a new class of double-faced white dwarfs. If more are found with similar temperatures, it will be a strong indication that their behaviour is connected to the rise in convective instabilities in the helium layer, and therefore double-faced white dwarfs could be a magnetic sub-sample of white dwarfs transitioning out of the DB gap. Thanks to current and future photometric surveys, like ZTF and the upcoming Vera Rubin Observatory, as well as to the new all-sky spectroscopic surveys (DESI, SDSS V, WEAVE), discovering such spectroscopically variable DB white dwarfs will soon become easy. Finding and studying a class of transitioning objects will help shed light on the still poorly understood physical mechanisms that underpin spectral evolution in white dwarfs.

\begin{table}
\centering
 \caption{\textbf{SED fitting parameters}. We list the input data (first two rows) and results (last two rows) of the fit shown in Fig.~\ref{fig:allfit}. For each filter, we list the AB magnitude at phase 0 (upper value) and the difference in magnitude between phase 0 and phase 0.5 obtained by fitting the light curve with a sinusoid as in Fig.~\ref{fig:lc} and Extended Data Fig.~\ref{fig:lc_swift} (lower value). We fit for both the interstellar extinction and parallax by imposing a prior from a dust map and from the Gaia eDR3 value respectively (see Methods), and we here list both the priors and the fitting results. }
\medskip
\begin{tabular}{cccc}
\hline
\makecell{$\rm{UVOT-}$$UVW2$} & \makecell{$\rm{HiPERCAM-}$$u_s$} & \makecell{$\rm{HiPERCAM-}$$g_s$} & \makecell{$\rm{HiPERCAM-}$$r_s$} \  \\
\makecell{$18.60\pm0.08$ \\ $0.46\pm0.08$}& \makecell{$19.40\pm0.05$ \\ $0.22\pm0.01$} & \makecell{$19.83\pm0.05$ \\ $0.185\pm0.004$} & \makecell{$20.18\pm0.05$ \\ $0.160\pm0.004$} \\
\hline
 \makecell{$\rm{HiPERCAM-}$$i_s$} & \makecell{$\rm{HiPERCAM-}$$z_s$} & \makecell{$\rm{E(B-V)}\rm{- prior}$} & \makecell{$\rm{Parallax}~\rm{[mas] - prior}$}\\
 \makecell{$20.59\pm0.05$ \\ $0.143\pm0.008$} & \makecell{$20.98\pm0.05$ \\ $0.12\pm0.02$} & $0.03\pm0.01$ & $2.5\pm0.4$ \\ 
\hline
 \makecell{$T_{\rm{eff}}~\rm{[K] - H~face}$} & \makecell{$T_{\rm{eff}}~\rm{[K] - He~face}$} & \makecell{$\rm{Radius}~\rm{[km]}$} & \makecell{$\rm{E(B-V)}\rm{- fit}$}\\ 
 $34,900^{+1,300}_{-1,500}$ & $36,700^{+1,300}_{-1,600}$ & $3,400^{+700}_{-600}$ & $0.037^{+0.008}_{-0.008}$ \\
\hline
\makecell{$\rm{Parallax}~\rm{[mas] - fit}$} & \makecell{$\rm{Mass - CO ~core}~\rm{[M]}_\odot$} & \makecell{$\rm{Mass - ONe ~core}~\rm{[M]}_\odot$} &\\ 
 $2.5^{+0.4}_{-0.4}$ & $1.27\pm0.06$ & $1.21\pm0.06$ & \\
\hline

\end{tabular}
\label{tab:fitpar}
\end{table}

\begin{methods}

\renewcommand{\tablename}{Extended Data Table}
\setcounter{table}{0} 
\renewcommand{\figurename}{Extended Data Figure}
\setcounter{figure}{0} 

\begin{table}
\centering
 \caption{\textbf{Janus parameters} The ephemeris $T_0$, Barycentric Modified Julian Date in barycentric dynamical time (BMJD$_{\rm{TBD}}$), corresponds to a maximum in the lightcurve.}
\medskip
\begin{tabular}{cccc}
\hline
\makecell{$\rm{Gaia~ID}$} & \makecell{$\rm{Parallax~[mas]}$} &\makecell{$\mu_{\rm{RA}}~\rm{[mas/yr]}$} & \makecell{$\mu_{\rm{DEC}}~\rm{[mas/yr]}$}\\ 
$1863529616173400576$ & $2.45\pm0.44$ & $3.5\pm0.4$ & $-7.0\pm0.4$ \\
\hline
\makecell{$T_0$ $[\rm{BMJD}_{\rm{TDB}}]$} & \makecell{$P~\rm{[s]}$} & & \\
$59401.34710\pm0.00003$  & $898.023233\pm0.000016$ &  & \\
\hline
\end{tabular}
\label{tab:par}
\end{table}

\subsection{Period Detection}
Janus was found during a search for periodic variability on and around the white dwarf cooling track with ZTF\cite{2019PASP..131a8002B,2019PASP..131g8001G,2020PASP..132c8001D,2019PASP..131a8003M}, which has already yielded several results, including finding numerous double white dwarf binaries\cite{Burdge2019a,Burdge2019b,2020ApJ...905...32B} and an extremely massive and magnetic white dwarf which is most likely the result of a white-dwarf merger\cite{2021Natur.595...39C} . The targets were selected using Pan-STARRS (PS1) source catalogue\cite{2016arXiv161205560C}, cross-matched with a white dwarf catalogue\cite{2019MNRAS.482.4570G}, after imposing a photometric colour selection of $(g-r)<0.2$~mag and $(r-i)<0.2$~mag. As the sensitivity of period finding depends strongly on the number of samples in the lightcurve, we limited the search to those targets for which 50 or more photometric $5 \sigma$ detections are available in the ZTF archival data. In order to maximise the number of epochs for each lightcurve, we combined data from multiple filters by computing the median magnitude in each filter, and shifting the $g$- and $i$-band so that their median magnitude matched the $r$-band data. We used a graphics processing unit (GPU) implementation of the conditional entropy period finding algorithm\cite{Graham2013}. We cross-matched our candidates with the Gaia DR2 catalogue\cite{2018A&A...616A...1G} and visually inspected the lightcurves of those objects that lie below the main white dwarf cooling track in the Gaia colour-magnitude diagram. Janus stood out because of the high-significance detection of its short period and its blue and faint location in the colour-magnitude diagram (see Fig.~\ref{fig:cmd}).
 
 \begin{figure}
    \centering
    \includegraphics[width=\textwidth]{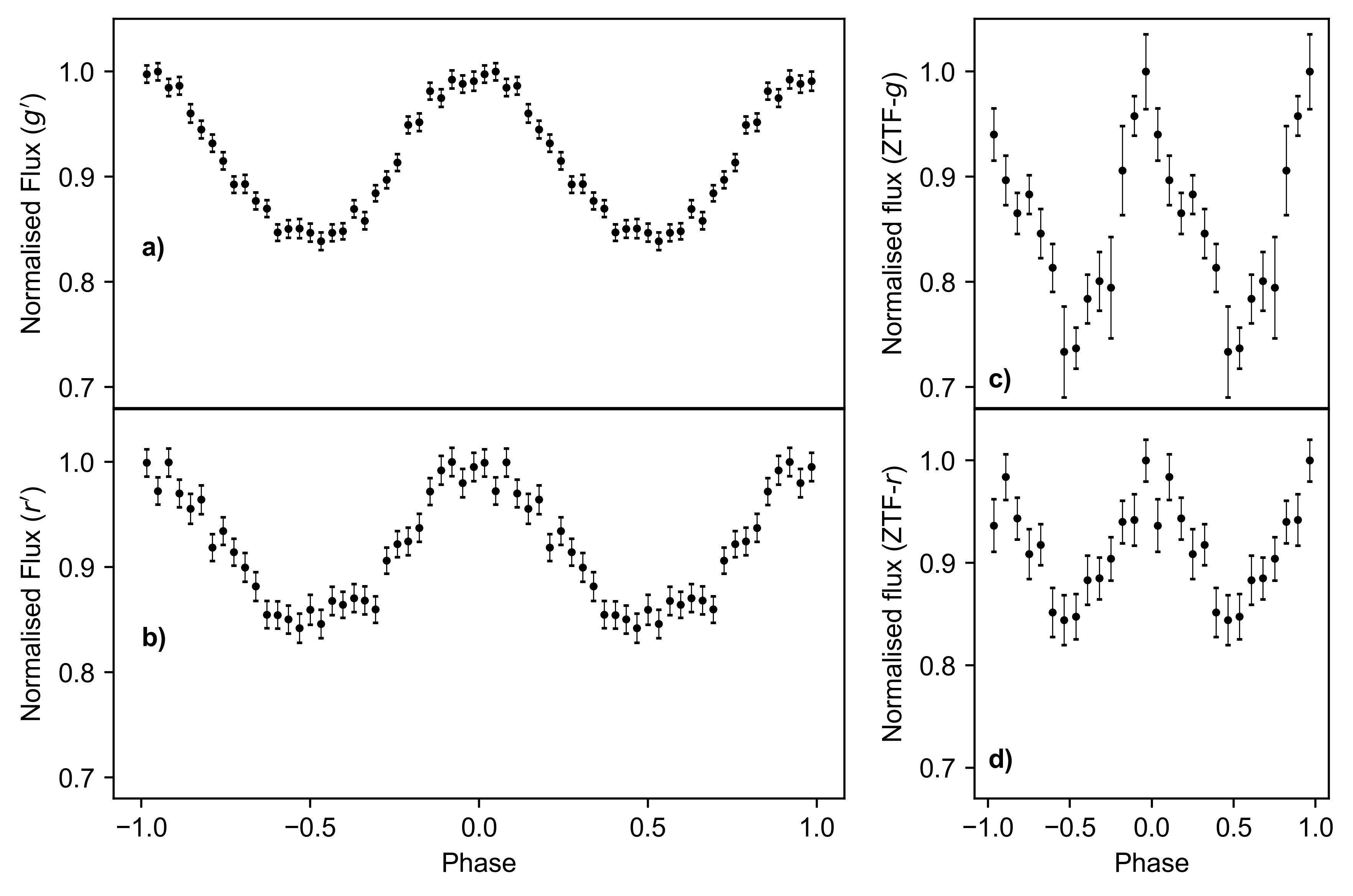}
    \caption{{\bf Janus ZTF and CHIMERA light curve} The left panels show the binned CHIMERA light curve phase-folded at a period of 14.97 minutes in the $g^\prime$-band (\textbf{a}) and in the $r^\prime$-band (\textbf{b}). The flux has been normalised to the maximum of the light curve in each band. The amplitude of the photometric variation is about 15\% peak-to-trough in both bands. The right panels show the similarly normalised ZTF discovery light curve in the ZTF $g$-band (\textbf{c}) and $r-$band (\textbf{d}). The error bars indicate 1$\sigma$ errors.}
    \label{fig:lc_ztf}
\end{figure}

\subsection{Keck LRIS observation}

We obtained the Keck LRIS spectra using the 600/4000 grism on the blue arm and the 600/7500 grating on the red arm. We used the $4\times 4$ binning on the blue channel and the $2\times 2$ on the red channel to reduce both readout noise and readout time. We reduced the Keck LRIS observations using a publicly available LPIPE automated data reduction pipeline\cite{2019PASP..131h4503P}.

\subsection{HiPERCAM observation}

The HiPERCAM data were obtained on two nights, Sep 6th and Sep 9th 2021, for a total of 2.1 hours. Both nights were affected by Sahara dust, which made obtaining an absolute flux calibration more challenging. We reduced the HiPERCAM data using the publicly available pipeline\cite{Dhillon2021}. The pipeline performed aperture photometry with a dynamic full-width-half-maximum, with variable aperture size equal to 1.7x the FWHM. We used $20$~s exposures in $u_s$, and $10$~s exposures in $g_{\mathrm{s}}$, $r_{\mathrm{s}}$, $i_{\mathrm{s}}$, and $z_{\mathrm{s}}$, and applied fringe map corrections the $z_{\mathrm{s}}$ band. The charge coupled devices (CCDs) were operated with conventional amplifiers using frame-transfer to effectively eliminate readout overheads. The extinction coefficients were derived using different observations during the night that spanned a large range in airmass. As the extinction was variable during both nights, we estimated our error in flux calibration from the range of extinction parameters obtained. The light curve is shown in Fig.~\ref{fig:lc} and the calibrated maxima and amplitude variation for each filter are listed in Table~\ref{tab:fitpar}.

\subsection{Swift UVOT observation}

The 5,459\,s \emph{Swift} observation (TOO proposal number 15866, target ID 14382, observation IDs 00014382001 and 00014355002) was equally divided in the UVOT filters UVW1, UVM2 and UVW2. We computed an average apparent magnitude of Janus in each filter using the uvotsource tool and a 5 arcsecond aperture centered on the source.  The values obtained are listed in Extended Data Table~\ref{tab:photometry}. We also requested a 5,388\,s observation in event mode to obtain a light curve in the UVW2 filter (TOO proposal number 17021, target ID 14382, observation ID 00014382004), which is shown in Extended Data Fig.~\ref{fig:lc_swift}. We find the same average magnitude as in the previous observation ($18.83\pm0.08$) and a peak-to-peak variation of $46\pm8\%$ (see also Table~\ref{tab:fitpar}).
 
 \begin{figure}
    \centering
    \includegraphics[width=0.6\textwidth]{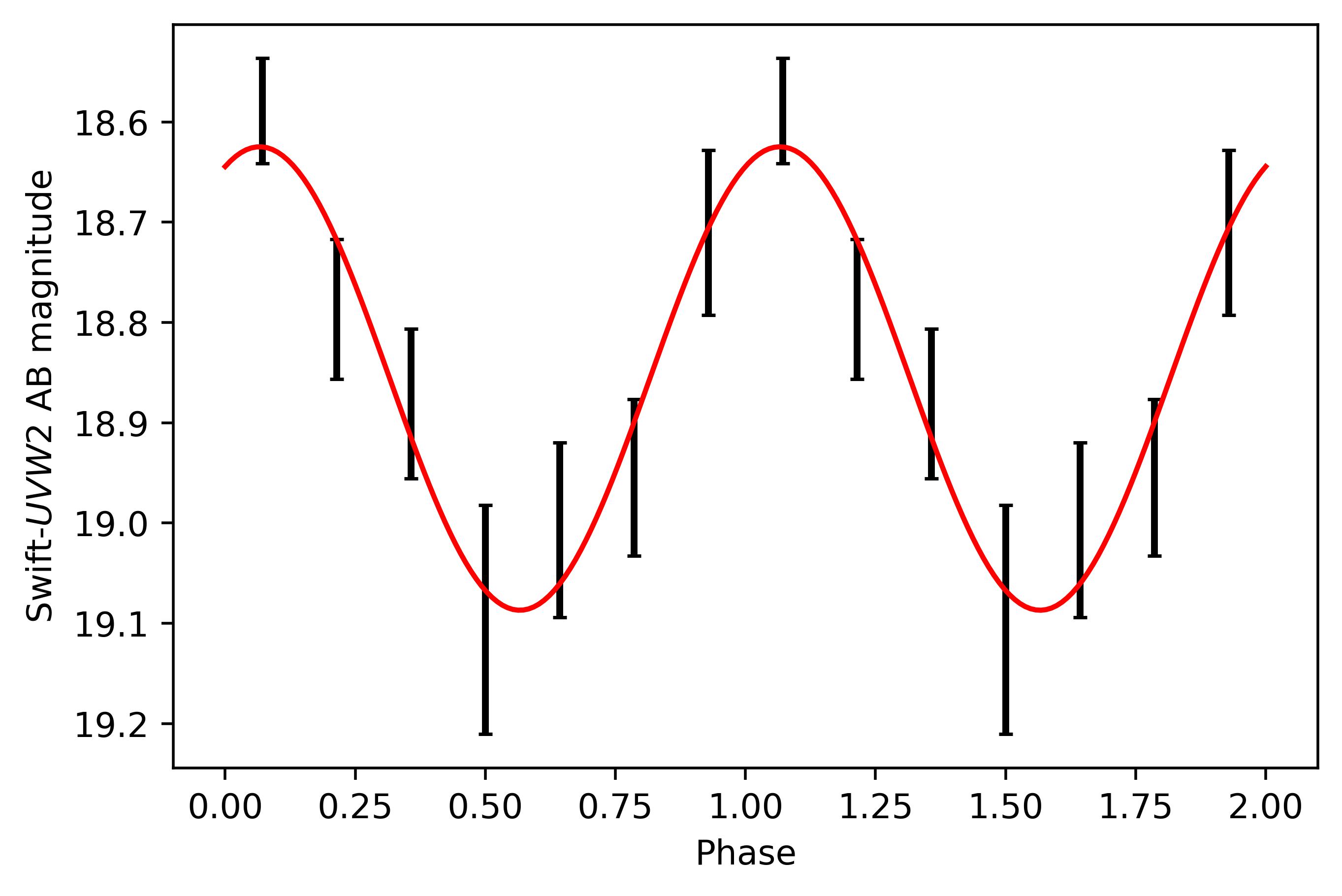}
    \caption{{\bf Janus Swift UVOT light curve} The plot shows the binned {\it Swift} UVOT light curve phase-folded at a period of 14.97 minutes in the UVW2 filter, the red solid line shows a sinusoidal fit with a peak-to-peak amplitude of $46\pm8\%$. The error bars indicate 1$\sigma$ errors.}
    \label{fig:lc_swift}
\end{figure}

 \begin{table}
\centering
 \caption{\textbf{Additional photometric data for Janus.} The values for the HiPERCAM and for the event-mode {\it Swift} observation (ID 00014382004) are given in Table~\ref{tab:fitpar}.}
\medskip
\begin{tabular}{cccc}
\hline
 \makecell{$\rm{PS1-}$$g$} & \makecell{$\rm{PS1-}$$r$}& \makecell{$\rm{PS1-}$$i$} & \makecell{$\rm{PS1-}$$z$} \\
 $19.87\pm0.03$ & $20.22\pm0.03$ & $20.61\pm0.02$ & $20.89\pm0.10$ \\
\hline
\makecell{$\rm{UVOT-}$$UVW2$} & \makecell{$\rm{UVOT-}$$UVM2$} &\makecell{$\rm{UVOT-}$$UVW1$} &  \\
$18.83\pm0.08$ & $19.00\pm0.08$ & $19.11\pm0.08$ &   \\
\hline
\end{tabular}
\label{tab:photometry}
\end{table}

\clearpage

\subsection{Photometric Fitting}
To determine the radius and effective temperature of Janus, we made use of the available Pan-STARRS\cite{2016arXiv161205560C} photometry and the Gaia\cite{2016A&A...595A...1G,2021A&A...649A...2L} parallax. In addition, we obtained \emph{Swift}\cite{2004ApJ...611.1005G} UVOT\cite{2005SSRv..120...95R} photometry (see previous section). The photometric data used in the fitting is listed in Extended Data Table~\ref{tab:photometry}. In order to estimate the temperature, radius, and reddening, we fitted the photometric data with the white dwarf 1-D model DA (hydrogen dominated) atmospheres of ref\cite{2011ApJ...730..128T} and with the 1-D model DB (helium dominated) atmospheres of ref\cite{2021MNRAS.501.5274C}. In order to account for extinction, we applied reddening corrections to the synthetic spectra using the Cardelli et al. 1989\cite{1989ApJ...345..245C} extinction curves, available at \texttt{https://www.stsci.edu/}. From the corrected models, we computed synthetic photometry using the \texttt{pyphot} package \\(\texttt{https://mfouesneau.github.io/docs/pyphot/}).

For the fit, we used a Levenberg-Marquardt algorithm, and the free parameters were the effective temperature $T_{\rm{eff}}$ and radius $R_*$ of the white dwarf as well as the interstellar reddening $E(B-V)$. For the reddening, we imposed a prior, based on the dust map by Green et al.\cite{2019ApJ...887...93G}, of $E(B-V)=0.03\pm 0.02$. The best fits for the hydrogen and helium models are shown in Extended Data Fig.~\ref{fig:phot}. In the fits, we assumed the nominal value of the Gaia parallax for the distance. The values obtained are $T_{\rm{eff}}=28,000\pm3,000$~K, $R_*=4,100\pm900$~km and $E(B-V)=0.03\pm 0.02$ for the hydrogen model and $T_{\rm{eff}}=36,000\pm4,000$~K, $R_*=3,600\pm800$~km and $E(B-V)=0.03\pm 0.02$ for the helium model.

\begin{figure}
    \centering
    \includegraphics[width=0.495\textwidth]{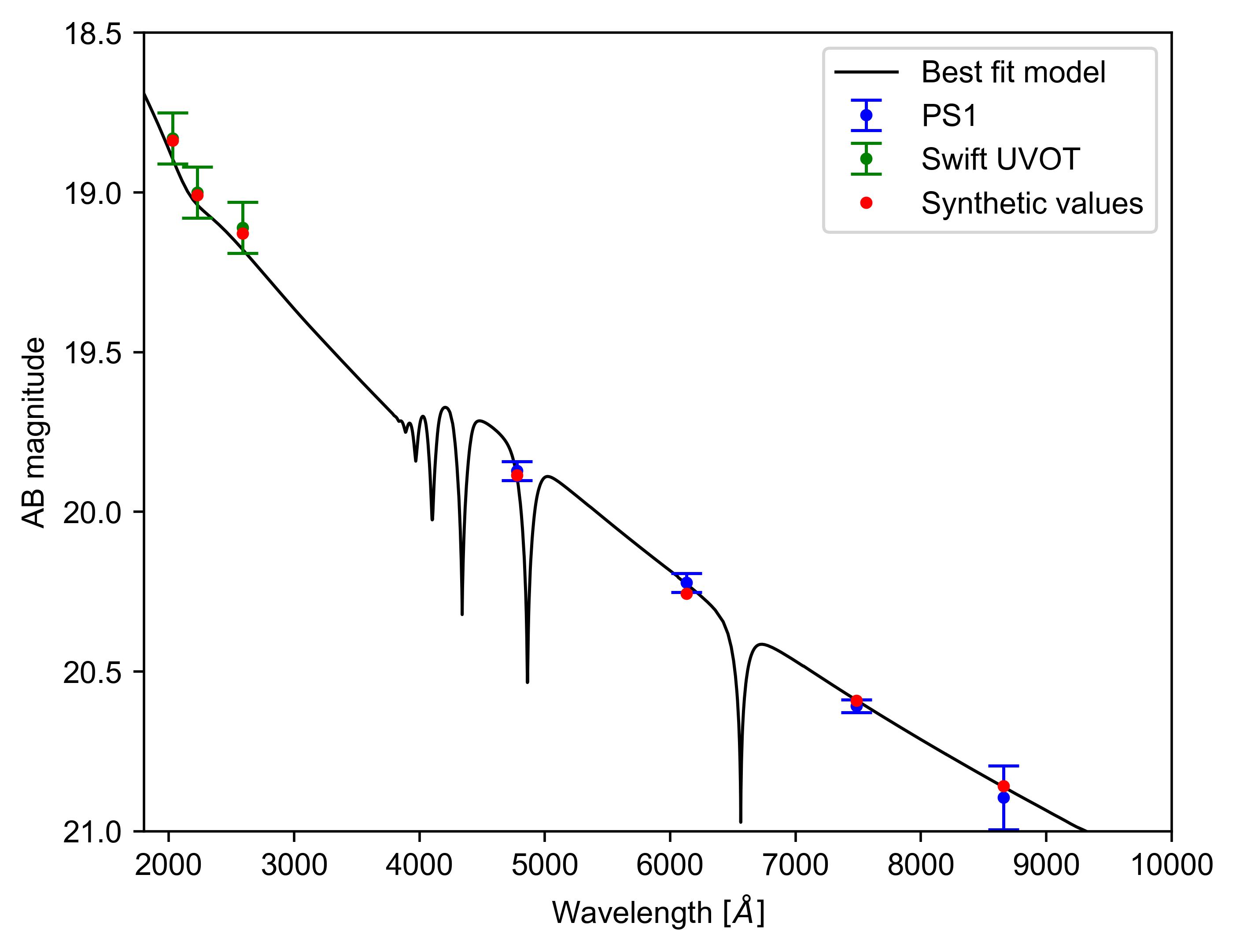}
    \includegraphics[width=0.495\textwidth]{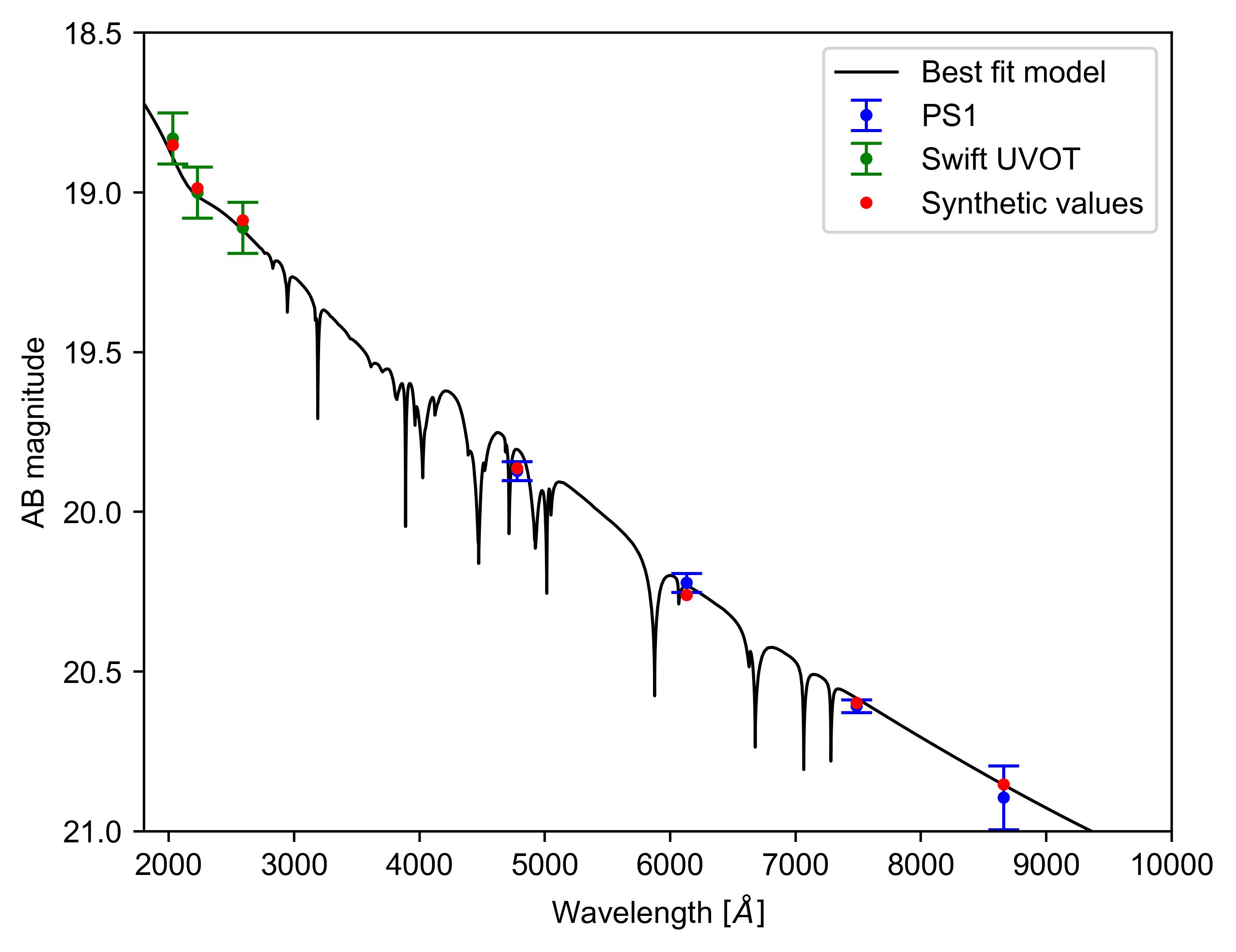}
    \caption{{\bf Photometric fit.} The black solid line shows the best-fitting model spectrum, fitted to Pan-STARRS and \emph{Swift} photometry to determine $T_{\rm{eff}}$, $R_*$ and $E(B-V)$. The synthetic photometric values (obtained from the black line) are shown in red, while the \emph{Swift} values are shown in green and the Pan-STARRS values in blue. Left: hydrogen atmosphere model; right: helium atmosphere model.} 
    \label{fig:phot}
\end{figure}

The SED changes with phase, as we can see from the HiPERCAM and \emph{Swift} data (Fig.s~\ref{fig:lc} and~\ref{fig:lc_swift}). Furthermore, the shape of the SED in the optical of a hydrogen-dominated white dwarf is quite different from a helium-dominated one at the same temperature. We therefore decided to simultaneously fit the SEDs of the brightest and the dimmest phases separately, employing the light curves obtained with HiPERCAM and \emph{Swift} UVOT UVW2. For the bright side (phase 0), we used the hydrogen models grid\cite{2011ApJ...730..128T}, while for the dimmest phase (phase 0.5) we used the helium models\cite{2021MNRAS.501.5274C}, trying several hydrogen-to-helium mass ratios ranging from 10$^{-30}$ (pure helium) to 10$^{-3}$ (a higher hydrogen content would cause hydrogen absorption features to appear in the spectrum at phase 0.5). We found that there was not an appreciable change in the fit by changing the hydrogen content in the DB models in the range considered, so we report only the fit with pure helium atmospheres for phase 0.5.  We used a Levenberg-Marquardt algorithm and the input data is listed in Table~\ref{tab:fitpar}: the HiPERCAM and \emph{Swift} UVW2 magnitudes at the brightest phase, the difference in magnitude in each filter from the brightest phase to the dimmest phase and a prior on reddening and distance that come from a dust map\cite{2019ApJ...887...93G} and from the Gaia eDR3 parallax measurement\cite{2021A&A...649A...2L}. The free parameters were the effective temperatures of the two faces and the radius of the white dwarf as well as the interstellar reddening and distance. The results from the fit are also listed in Table~\ref{tab:fitpar}.

To estimate the mass of the white dwarf, we compared the radius and temperatures obtained with white dwarf cooling models by interpolating publicly available tables\cite{2020ApJ...901...93B,2019A&A...625A..87C}. The models by Bedard et al.\cite{2020ApJ...901...93B} assume a carbon-oxygen core composition, and, at a radius of $3,400^{+700}_{-600}$~km predict a mass of $1.27\pm0.06$ M$_\odot$ for a DA white dwarf with $T_{\rm{eff}}=34,900^{+1,300}_{-1,500}$~K, and a mass of $\sim1.26\pm0.06$ M$_\odot$ for a DB white dwarf with a temperature $T_{\rm{eff}}=36,700^{+1,300}_{-1,500}$~K. For oxygen-neon core composition, we used the models by Camisassa et al.\cite{2019A&A...625A..87C}, and we obtained
a mass of $\sim1.21\pm0.05$ M$_\odot$ for a DA white dwarf with $T_{\rm{eff}}=34,900^{+1,300}_{-1,500}$~K, and a mass of $\sim1.21\pm0.06$ M$_\odot$ for a DB white dwarf with a temperature $T_{\rm{eff}}=36,700^{+1,300}_{-1,500}$.

\subsection{Magnetic field strength} 
Magnetic fields can be detected in the spectra of white dwarfs because of Zeeman splitting. At field strengths between about 1 and 5 MG, in the linear Zeeman regime, an absorption line is split into an unshifted central component, a redshifted component and a blueshifted component, and the separation in energy between the components varies linearly with the field strength\cite{2015SSRv..191..111F}. At higher magnetic fields, the central component is also blueshifted. For a white dwarf with lines as broad and weak as Janus, the Zeeman splitting is unresolvable below $\sim1$~MG for most magnetic field geometries. Depending on the geometry of the field, Zeeman splitting can be unresolved at higher fields as well, as is observed in WD J0103-0522\cite{2020MNRAS.497..130T}. WD J0103-0522 is an ultra-massive white dwarf with strongly diluted Balmer lines that does not show any Zeeman splitting in the spectrum; the Balmer lines, however, are blueshifted, possibly indicating a magnetic field of the order of 5~MG (although alternative explanations have been proposed \cite{2022A&A...659A.157S}). As Janus does not show any obvious splitting nor shifts in the hydrogen or helium lines in its spectrum, we can infer that, if a magnetic field is present, it has to be weaker than a few MG.

\subsection{White Dwarf Models}

The double-faced nature of Janus is intriguing, especially because the temperature of the white dwarf might imply a connection between its behaviour and the spectral transition between DA and DB white dwarfs. In the main text, we presented some models that could explain the spectroscopic and photometric variability of the white dwarf; we here analyse in more detail the constraints and shortcomings of each model, as well as present other possibilities that we consider less likely.

\noindent {\it The dilution model.} At a temperature of about 35,000~K, the convective instabilities in the helium layer are quite marginal and limited to a very small depth range; a small change in pressure due to the presence of a magnetic field can easily suppress convection and lead to a fully radiative structure. This is unlike cooler white dwarfs where the adiabatic gradient is significantly smaller than the radiative gradient, therefore making it essentially impossible to completely kill convective instabilities. The spectrum of Janus does not show any Zeeman splitting or wavelength shift, and therefore we can infer that the magnetic field, if present, has to be lower than a few MG. A careful calculation of the field needed to stop convection with a realistic magnetic field structure in a stratified atmosphere would require MHD simulations that are beyond the scope of this paper. However, we can get an estimate of the minimum field strength needed to suppress convection by assuming that the magnetic pressure equals the gas pressure ($\beta = P_{\rm gas}/P_{\rm mag} = 1$) at an optical depth of about one, which would be enough to make hydrogen float back to the surface. In a nearly pure helium atmosphere\cite{2021MNRAS.501.5274C} (H/He$=10^{-10}$), the gas pressure at $\tau_R \sim 1$ is of the order $P_{\rm{gas}}=10^{7}$~erg~cm${^{-3}}$, which would imply a magnetic field B $\approx16$ kG. Of course, this is just a ballpark estimate and the threshold magnetic field could easily be an order of magnitude higher, assuming that upward overshoot plumes can still mix the photosphere through convective dilution at $\beta \sim 0.01$. In order to explain the double-faced nature of Janus, the magnetic field on the surface would have to be greater than this value on the hydrogen side and lower on the helium side.  

In this dilution model, the thickness of the hydrogen layer is also constrained. From the phase-resolved spectroscopy, we know that a large fraction of one face is covered by optically thick hydrogen, because we do not see absorption lines from any other element, and this means that the hydrogen content must be of the order of $10^{-17}$~M$_\odot$ or larger (assuming a $\log g$ of 9 and a temperature of $\sim30,000$~K). On the other hand, we do not see hydrogen on the mixed side.
Given the size of the helium convection zone\cite{2019MNRAS.490.1010C} and assuming two pressure scale heights of convective overshoot as well as a surface composition of H/He = $10^{-2}$, the hydrogen content should not exceed $\sim10^{-15}$~M$_\odot$. Otherwise, even when diluted, hydrogen would still show in the spectrum on the helium side.
Thus, in this scenario, both the amount of hydrogen and the magnetic field strength have to be somewhat fine-tuned, and Janus would have to be a rather rare type of white dwarf. Since the white dwarf is quite distant, at more than 400 pc from Earth, this is not a problem per se, but it would be hard to justify if more objects of the same type were to be found.

\begin{figure}
    \centering
    \includegraphics[width=0.9\textwidth]{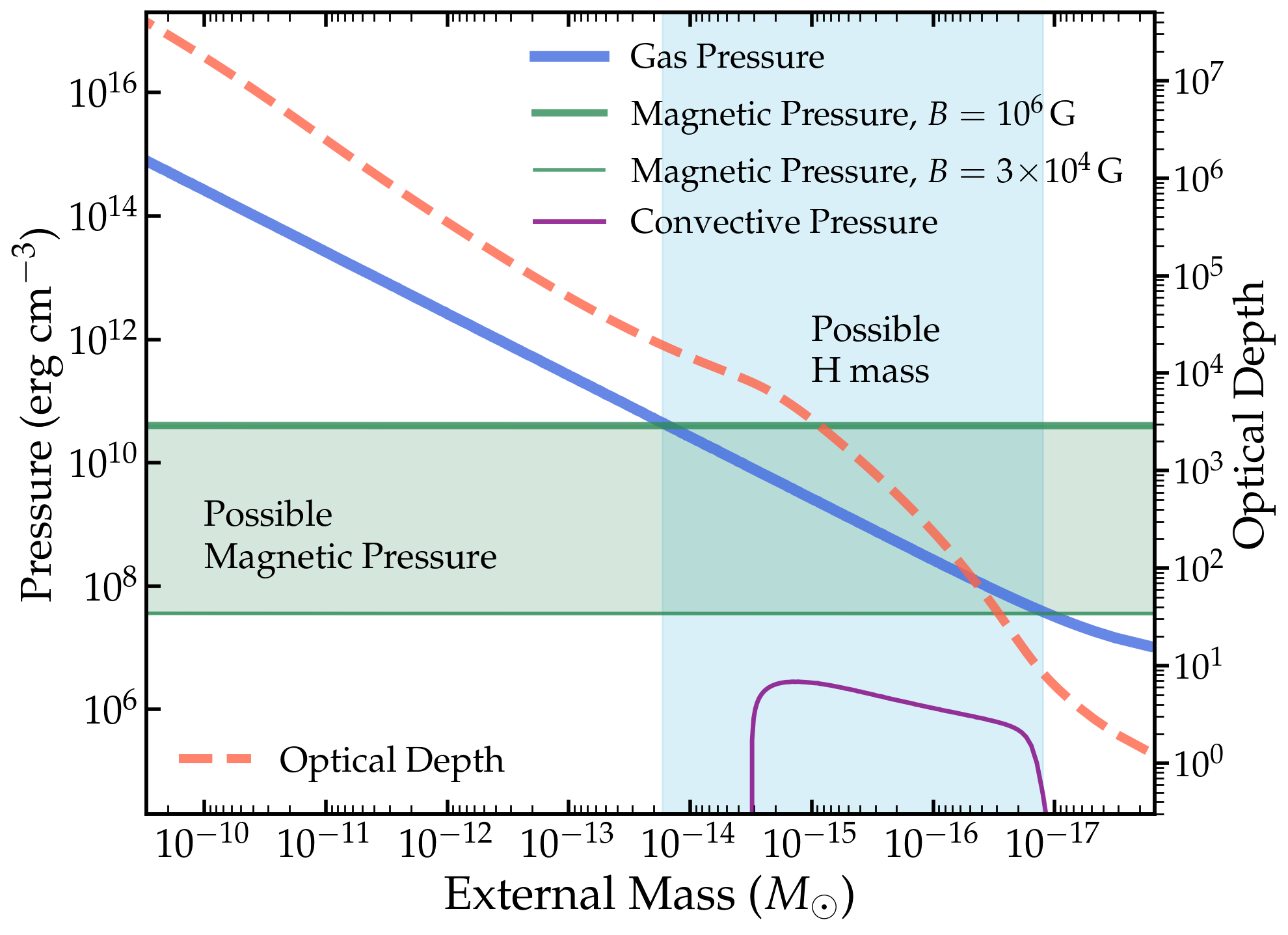}
    \caption{{\bf Pressure inside the white dwarf.} The pressure from gas, magnetic field, and convective motions as a function of depth within a white dwarf of 1.259 solar masses with a ONe core and a pure helium atmosphere. The x-axis indicates the exterior mass (the photosphere is on the right), and the right-hand y-axis indicates the optical depth. The shaded blue vertical strip indicates an approximate plausible range of H masses, while the shaded green horizontal strip is a plausible range of magnetic field pressure. }
    \label{fig:hd}
\end{figure}

\noindent {\it The ocean model.} Another possibility that we mentioned in the main text is the fact that hydrogen might diffuse toward a higher-field region. In order to test this hypothesis, we generated white dwarf models using the MESA stellar evolution code\cite{Paxton2018}. Beginning with MESA's $1.259 \, M_\odot$ oxygen-neon white dwarf model, we stripped the hydrogen using \verb|remove_H_wind_mdot| until no hydrogen remained. We then turned on diffusion and gravitational settling and evolved the star to a surface temperature of $31,000 \, {\rm K}$. Figure \ref{fig:hd} shows a pressure profile of the stellar atmosphere as a function of external mass $M_{\rm ext}$.  We also plot the corresponding magnetic pressure (not included in the MESA model) for a dipole field $B = B_0 (r/R)^{-3}$ of strengths that range from $B_0 = 30 \, {\rm kG}$ for which the magnetic pressure $P_{\rm mag} = B^2/(8 \pi)$ equals the gas pressure at an optical depth of $\sim10$, to $B_0 = 1 \, {\rm MG}$, which is close to the limit we get from the observed spectrum. In the MESA model, the gas pressure reaches $10^7$~erg~cm${^{-3}}$ at an optical depth $\tau_R \sim 3$ instead of  $\tau_R\sim1$, as in the atmospheric models employed above; this discrepancy is due to the fact that the photospheric structure of the MESA model is only approximate.

\begin{figure}
    \centering
    \includegraphics[width=0.8\textwidth]{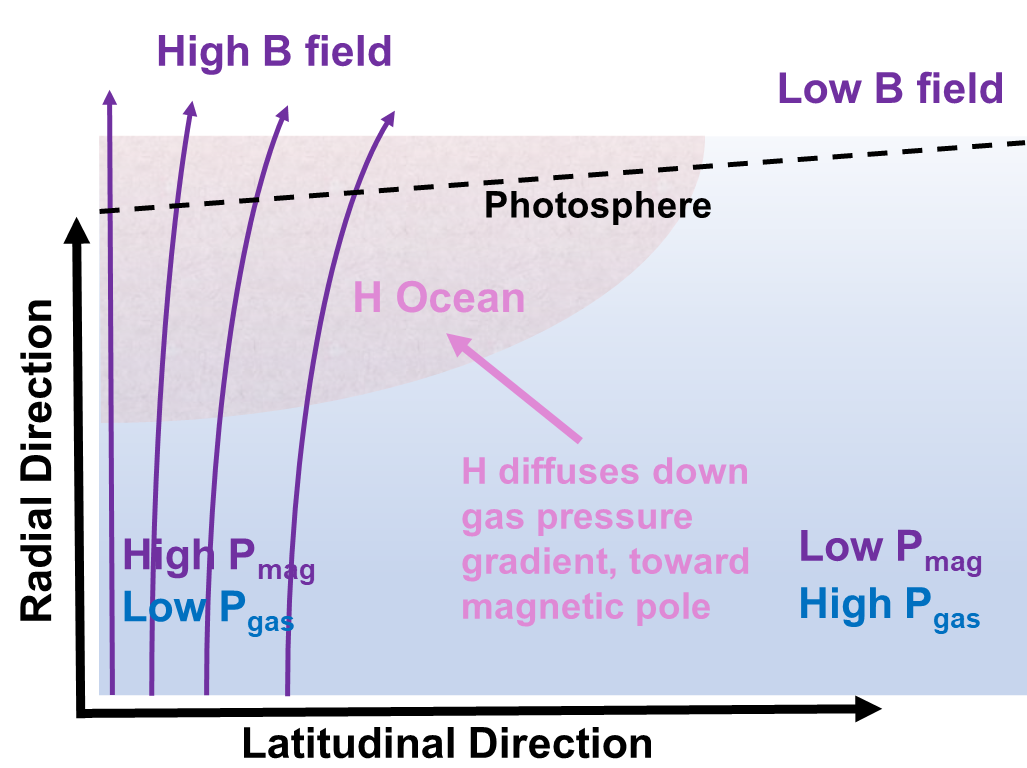}
    \caption{{\bf Scheme of the hydrogen diffusion scenario in the presence of a magnetic field.} Hydrogen diffuses upwards (due to its low mass) and toward the magnetic pole (due to the ion pressure gradient and its low charge). If there is sufficiently little hydrogen within the white dwarf, it will only cover the photospheric layers near the magnetic pole, possibly explaining the variable surface composition of Janus.} 
    \label{fig:hf}
\end{figure}

If convection is prevented in the near-surface layers, then separation of hydrogen and helium may occur via diffusion and gravitational settling. Consider hydrogen as a trace pollutant in a helium atmosphere. Then the diffusion velocity of hydrogen relative to helium is \cite{bauer:19}
\begin{equation}
    \vec{v}_{\rm diff} = D \bigg[-\frac{\nabla c_H}{c_H} - \bigg(\frac{Z_{\rm H}}{Z_{\rm He}} A_{\rm He} - A_{\rm H} \bigg)\frac{m_p \vec{g}}{k T} + \bigg(\frac{Z_{\rm H}}{Z_{\rm He}} -1 \bigg) \frac{\nabla p_{\rm ion}}{p_{\rm ion}} + \alpha_T \frac{\nabla T}{T} \,, \bigg]
\end{equation}
where $c_H=n_H/(n_H+n_{He})$  is the concentration of hydrogen, $p_{\rm ion}$ is the ion pressure, $\alpha_T$ is the thermal diffusion coefficient, and A and Z refer to the mass and charge of each species.
Consider the fully ionized case where $Z_{\rm H}=1$, $Z_{\rm He}=2$, $A_{\rm H}=1$, and $A_{\rm He}=4$. Then the second term will cause hydrogen to rise upwards (against the direction of gravity), as we expect in a white dwarf. The third term will cause hydrogen to diffuse against the ion pressure gradient, i.e., towards regions of low ion pressure and hence low gas pressure. Hence, we expect hydrogen atoms to diffuse towards the magnetic poles where the gas pressure is smaller.

Hydrogen could thus form an ``ocean" at the surface of the white dwarf, with a deeper ocean at the magnetic poles due to the diffusion described above (Fig. \ref{fig:hf}). The lateral diffusion of hydrogen toward the magnetic pole will be efficient in layers where $P_{\rm mag} \gtrsim P_{\rm gas}$ such that there is a large gas pressure gradient due to the magnetic field. An estimate for the upper limit to the amount of hydrogen in Janus is the mass coordinate (Fig. \ref{fig:hd}) at which $P_{\rm mag} \sim P_{\rm gas}$. For much larger hydrogen masses, the ocean basin where $P_{\rm gas} \lesssim P_{\rm mag}$ will ``fill up" and the hydrogen ocean will cover the entire photosphere. If the hydrogen mass is smaller than the mass contained within the final scale height of the star, it will not be optically thick enough to create strong Balmer lines, setting an approximate lower limit for the amount of hydrogen present. From the range of magnetic fields considered, we obtain a range of hydrogen masses between $\sim10^{-17}$ solar masses to $\sim10^{-14}$ solar masses (Fig. \ref{fig:hd}).
More detailed multi-dimensional diffusion models will be needed to test this scenario and provide more quantitative estimates of the magnetic field strength and hydrogen content of white dwarfs like Janus.

\noindent {\it DA+DB binary.} Several white dwarfs that show hydrogen-dominated atmospheres and traces of helium, originally classified as DAB, have been later identified as binaries containing a hydrogen-rich (DA) and a helium-rich (DB) white dwarf\cite{1994ApJ...429..369W,2002MNRAS.334..833M,2009ApJ...696.1461L}. However, as we explain in the main text, a binary nature can be excluded in the case of Janus because we do not detect Doppler shifts in the spectrum, which should be apparent at a 15 minute period, and because  the light curve is sinusoidal and does not show eclipses. 

\noindent {\it Pulsations.} Both DA and DB white dwarfs with temperatures close to that of Janus have been observed to pulsate\cite{2017ApJ...835..277H,2020MNRAS.497L..24R}. At these temperatures (or slightly colder ones), two mechanisms are known that can drive pulsations: on one hand, the V777 Her (or DBV) instability strip lays between $\sim20,000$ and $\sim30,000$~K and is caused by partial He ionisation in the atmosphere of DB white dwarfs; on the other hand, it has been suggested that, in transitioning white dwarfs at the red end of the DB gap, pulsational instabilities can be driven by radiative heat exchange in the superadiabatic atmospheric layer\cite{2007AIPC..948...35S,2008MNRAS.389.1771K}. As we already stated in the main text however, the spectral changes observed in Janus cannot be explained simply by the changes in temperature caused by pulsations; furthermore, we do not see any indication of pulsations: the rotation period has been stable for three years and no other significant periods can be observed in the periodogram.

\subsection{Weakness of the absorption lines}
As we mention in the main text, the absorption features on both faces appear weak for the temperature that we infer from the SED. If we compare the spectrum of the hydrogen face with our hydrogen models for example, the absorption lines are as weak as in models with temperatures around $50,000$~K (see Extended Data Fig.~\ref{fig:compspec}, upper panel). Our helium models' grid only goes up to $40,000$~K and in our highest temperature models, the lines are still stronger than in the observed spectrum at phase 0.5. However, at $40,000$~K we already see a He II absorption line at 4,685 \AA, which is even stronger at higher temperatures and which is absent in Janus' spectrum. For this reason, we assume that the temperature of Janus is actually close to what we infer from the SED and that the lines are weakened by some other mechanism.

In the main text, we propose a few possible mechanisms to explain the double-faced nature of Janus and, in all cases, we invoke the presence of a small magnetic field threading the surface of the white dwarf. If the presence of a magnetic field can also modify the thermal structure of the atmosphere, as for example by flattening the temperature gradient, it could weaken the absorption lines, although such an effect remains to be demonstrated. Tremblay et al.\cite{tremblay:15} have shown that magnetic fields of a few kG can significantly impact the thermal stratification in the upper layers of convective DA white dwarfs, although not enough to weaken the absorption lines as observed in Janus. Additionally, observations of the cool, metal-polluted, and magnetized white dwarf LHS 2534 show that the magnetic field can affect the atmospheric structure, exacerbating the narrowness of weaker metal lines\cite{2021NatAs...5..451H}. More calculations are needed that fully account for the influence of magnetic fields on the hydrostatic equilibrium of a white dwarf's atmosphere at different field strengths and atmospheric compositions.

\begin{figure}
    \centering
    \includegraphics[width=0.85\textwidth]{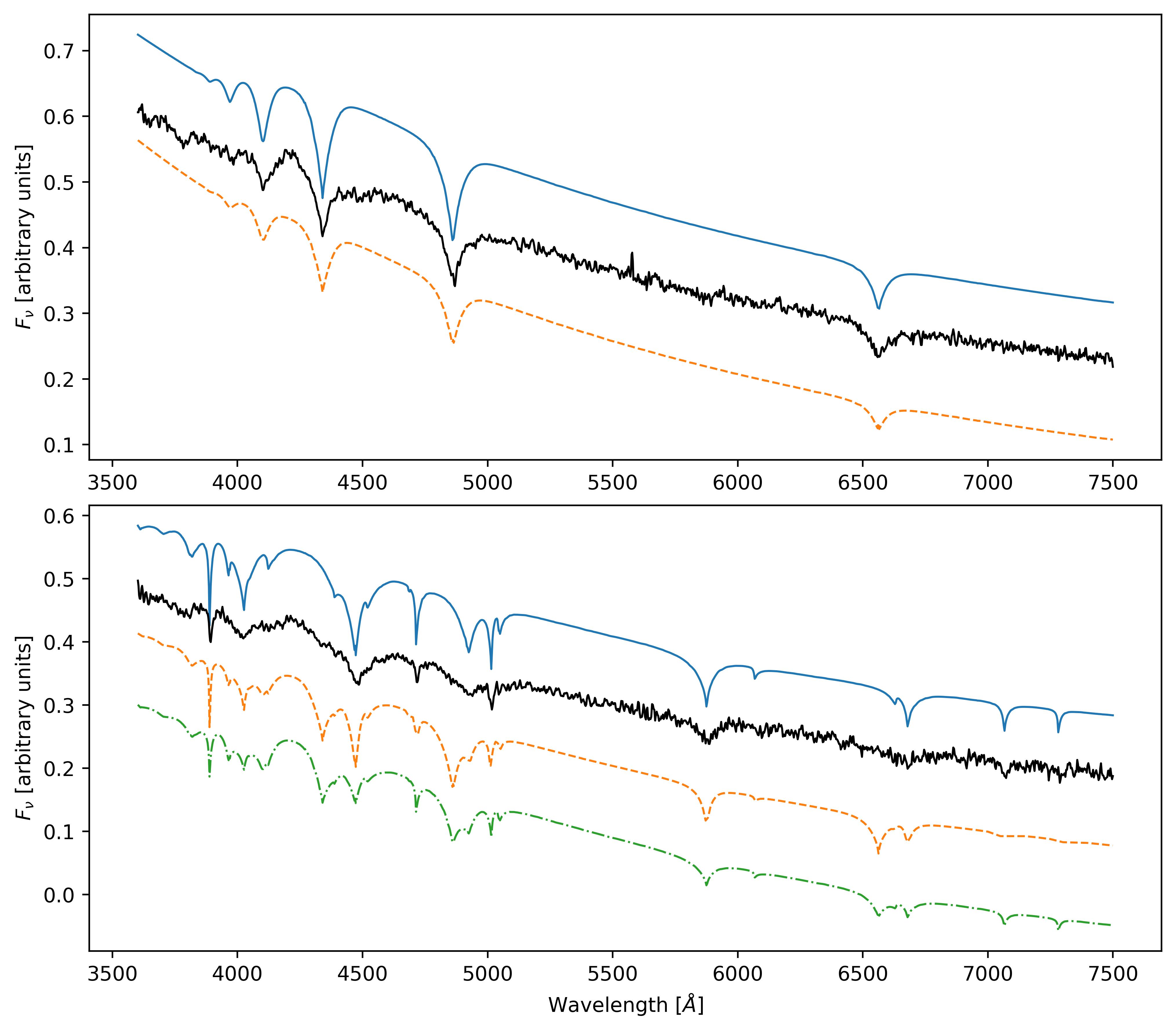}
    \caption{{\bf Comparison of phase-resolved spectra with models.}  {\bf Upper panel}: a comparison of the spectrum at phase~$0$ (the hydrogen phase, shown in black in the middle) with two synthetic models: on top in blue is a pure hydrogen model\cite{2011ApJ...730..128T} at $34,900$~K (the temperature inferred for the hydrogen phase from the SED) and at the bottom a pure hydrogen model at $50,000$~K. The lines in the observed spectrum are too weak compared to the model at $34,900$~K and resemble the weak lines of the model at $50,000$~K. {\bf Lower panel} from top to bottom: a pure helium model\cite{2021MNRAS.501.5274C} in blue, the observed spectrum at phase~$0.5$ in black (the helium phase), a homogeneously mixed atmosphere model with a hydrogen-to-helium mass ratio of 1 in orange\cite{2021MNRAS.501.5274C}, and a composite atmosphere in green in which 50\% of the flux is from a pure DA atmosphere\cite{2011ApJ...730..128T} and 50\% from a DB\cite{2021MNRAS.501.5274C}. The models are calculated for a temperature of $36,700$~K  (the temperature inferred for the helium phase from the SED). Neither a mixed nor a composite atmosphere can explain the weakness of both the hydrogen and helium lines.  } 
    \label{fig:compspec}
\end{figure}

The strength of the lines can be reduced if the atmosphere has a mixed composition; however, if we increase for example the hydrogen content in our models so that in the mixed atmosphere spectrum the helium lines appear as weak as in the observed spectrum at phase~$0.5$, the model spectrum also shows strong hydrogen lines, which we do not see in the observed spectrum. The same goes for the hydrogen face. A mixed composition is therefore insufficient to explain the weakness of the lines. Extended Data Fig.~\ref{fig:compspec} shows a mixed atmosphere model\cite{2021MNRAS.501.5274C} at $36,700$~K (the temperature inferred for the helium phase from the SED, Table~\ref{tab:fitpar}) with equal abundances of hydrogen and helium (orange dashed line): strong hydrogen lines are present and still the helium lines are too strong compared to the lines in the observed spectrum. 

The double-faced nature of Janus tells us that the composition of the white dwarf's surface is not homogeneous, and therefore even at phase $0$ and at phase $0.5$, when we see lines of only hydrogen or helium, a small fraction of the surface facing the line of sight might still be dominated by the other element and its emission can dilute the absorption lines. However, also in this case, in order for the lines to be weakened as much as in the observed spectrum, the lines of the other element would have to appear in the spectrum as well. In the lower panel of Extended Data Fig.~\ref{fig:compspec}, we show a simple example in which half of the surface is covered by pure hydrogen\cite{2011ApJ...730..128T} and half of the surface is covered by pure helium\cite{2021MNRAS.501.5274C} (green dashed-and-dotted line): the helium lines appear almost as weak as in the observed spectrum at phase~$0.5$, but both hydrogen and helium lines appear in the synthetic spectrum. The only way to dilute the lines at such temperature without other lines appearing in the spectrum would be with a featureless (DC) component. 

Extended Data Fig.~\ref{fig:DC} shows a fit to the lines at phase~$0$ and phase~$0.5$: the models were obtained by combining pure hydrogen\cite{2011ApJ...730..128T} and pure helium\cite{2021MNRAS.501.5274C} atmosphere models with featureless black bodies. The temperature was set to the one obtained from the SED fitting ($34,900$~K for phase~$0$ and $36,700$~K for phase~$0.5$, Table~\ref{tab:fitpar}) to both the atmosphere and the black body, while the fraction of surface emitting as a black body and the surface gravity were left as free parameters. Our line-fitting method is similar to the routine outlined in Liebert et al. 2005\cite{2005ApJS..156...47L}: we fit the spectrum with a grid of spectroscopic models combined with a polynomial in $\lambda$ (up to $\lambda^9$) to account for calibration errors in the continuum; we then normalise the spectrum using this smooth function picking normal points at a fixed distance in wavelength to the centres of the lines and finally use our grid of model spectra combined with black body emission to fit the lines and extract the values of the fraction of surface emitting as a black body and the logarithm of the surface gravity ($\log \rm{g}$). The nonlinear least-squares minimisation method of Levenberg-Marquardt is used in all our fits. In order to explain the weakness of the lines, about 40\% of the flux would have to be featureless.

\begin{figure}
    \centering
    \includegraphics[width=0.7\textwidth]{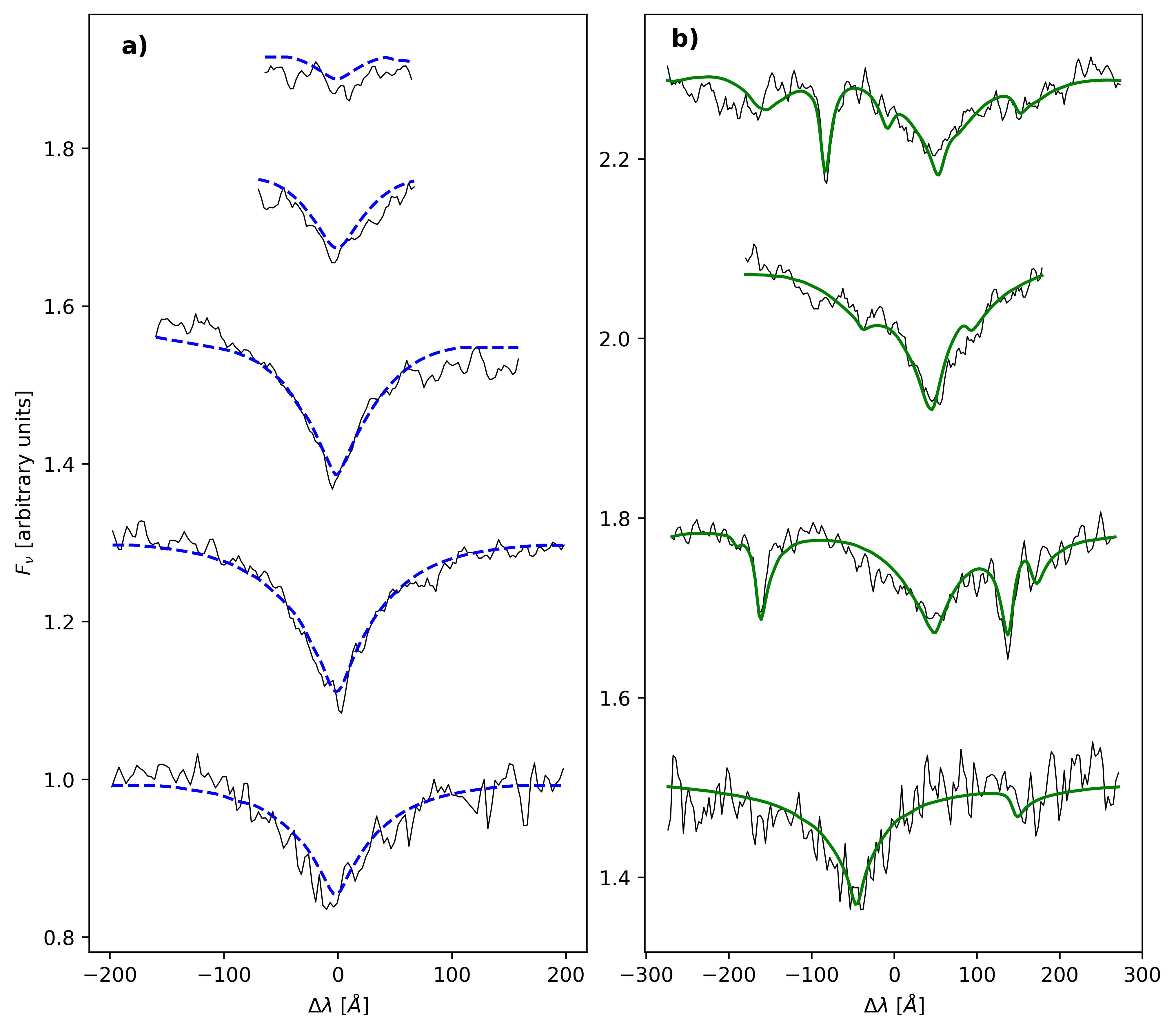}
    \caption{{\bf Fitting of the lines with a featureless component.} {\bf a)} An atmospheric model fit of the Balmer lines at phase~$0$ (in black), from bottom to top: H$\alpha$, H$\beta$, H$\gamma$, H$\delta$, H$\epsilon$. The blue dashed line shows the best-fitting model: $T_{\rm{eff}}=34,900$~K, $\log g = 9.1$ and fraction of surface emitting as a black body: 36\% . {\bf b)} Fitting of the He I absorption lines at phase~$0.5$ (in black). The $0$ on the x-axis corresponds to, from bottom to top: 5925~\AA, 4880~\AA, 4430~\AA\ and 3975~\AA. The green line shows the best-fitting model:$T_{\rm{eff}}=36,700$~K, $\log g = 9.1$ and fraction of surface emitting as a black body: 40\%} 
    \label{fig:DC}
\end{figure}

The weakness of the lines could be explained if Janus was in a binary with a highly magnetised white dwarf. Highly magnetised white dwarfs often show featureless spectra, due to strong magnetic broadening, and if Janus was in a binary with such a white dwarf, the weakness of the lines in its spectrum could be explained by contamination from the featureless spectrum of the companion. Since we do not detect any Doppler shifts in the lines, the companion would have to be in a large orbit (at least a few hours period). Additionally, as all the lines are weakened in a similar fashion, the temperature of the companion would have to be similar to that of Janus. If we assume the radii of the two objects to be also similar, they would then contribute equally to the observed SED and therefore the radii of both objects would be of the order of $R_*/\sqrt{2}\sim2,400$~km, where $R_*$ is the radius that we inferred from the SED fitting. This is close to the smallest radius ever observed for a white dwarf\cite{2021Natur.595...39C}, and would imply that both objects have masses above $1.3$~solar masses. Such massive white dwarfs are rare, and therefore we find it unlikely that the first ever observed double-faced white dwarf happens to be a 1.3~M$_\odot$ white dwarf in a binary with another 1.3~M$_\odot$, highly magnetised white dwarf.

\subsection{Other possible candidates}
The strength of the helium absorption lines in the spectrum of PG 1210+533, a DAO white dwarf at the blue end of the DB gap, has been observed to vary over decades\cite{1994ApJ...432..305B,2010ApJ...720..581G}. Gianninas et al.\cite{2010ApJ...720..581G} suggested that the variability in the white dwarf's spectrum could be caused by the onset of diffusive equilibrium after the extinguishing of the stellar winds, i.e. by the fact that the white dwarf might be entering the DB gap (the temperature of the white dwarf is $\sim45,000$~K). More follow-up is required to understand if the variation is periodic and if the variability in PG 1210+533 and in Janus might be related.

Recently, an ultra-massive DBA white dwarf was discovered to show photometric variability over a short period of 353.456~s, WD 1832+089\cite{2020MNRAS.499L..21P}. The phase-average spectrum shows helium lines and very weak hydrogen lines, and the temperature of the white dwarf is very similar to that of Janus, $\sim35,000$~K. WD 1832+089 is therefore a strong candidate to be of the same class as Janus and GD 323, but phase-resolved spectroscopy is needed to verify it.

The highly magnetised DBA white dwarf Feige~7 could also be a member of this class. Analysis of the phase-resolved intensity and circular polarization spectra for the object have revealed variability in both magnetic field strength and surface composition; in particular, Achilleos et al.\cite{1992ApJ...396..273A} invoked a complex surface morphology of areas of different mixed compositions of hydrogen and helium as well as patches of pure H and pure He to explain the spectral variations. The white dwarf has a magnetic field as high as 57~MG at the poles and it is colder than Janus ($\sim20,000$~K); if the cause of the varying surface composition is an interplay between magnetic field and convection, the colder temperature could be related to the much higher field strength.

\end{methods}

\let\oldthebibliography=\thebibliography
\let\oldendthebibliography=\endthebibliography
\renewenvironment{thebibliography}[1]{%
     \oldthebibliography{#1}%
     \setcounter{enumiv}{ 33 }%
}{\oldendthebibliography}

\begin{addendum}
     \item[Data Availability] Upon request, I.C. will provide the reduced photometric lightcurves and spectroscopic data, and available ZTF data for the object. The spectroscopic data and the optical photometric lightcurves are also available in the GitHub repository \texttt{https://github.com/ilac/Janus}, while ZTF data is accessible in the ZTF database. The astrometric data from Gaia and photometric data from Gaia, Pan-STARSS, and \emph{Swift} are already in the public domain, and they are readily accessible in the Gaia and Pan-STARSS catalogues and in the \emph{Swift} database.
 
 \item[Code availability] We used \texttt{astropy}\cite{2022ApJ...935..167A}, the  \texttt{pyphot} package \\ (\texttt{https://mfouesneau.github.io/docs/pyphot/}) and the \texttt{corner.py} package\cite{corner}. The LRIS spectra were reduced using the \texttt{Lpipe} pipeline\cite{2019PASP..131h4503P}. Upon request, I.C. will provide the code used to analyse the spectroscopic and photometric data.
 
\end{addendum}

\begin{addendum}
    \item The authors would like to dedicate this work to the memory of their good friend and colleague Thomas Marsh. The authors thank Dimitri Veras and Tim Cunningham for insightful discussions. I.C. thanks the Burke Institute at Caltech for supporting her research. P.-E.T. received funding from the European Research Council under the European Union’s Horizon 2020 research and innovation programme number 101002408 (MOS100PC), the Leverhulme Trust Grant (ID RPG-2020-366) and the UK STFC consolidated grant ST/T000406/1. T.R.M. and I.P. were funded STFC grant ST/T000406/1. S.G.P. acknowledges the support of a STFC Ernest Rutherford Fellowship. This research was supported in part by the National Science Foundation under Grant No. NSF PHY-1748958. This work is based on observations obtained with the Samuel Oschin Telescope 48-inch and the 60-inch Telescope at the Palomar Observatory as part of the Zwicky Transient Facility project. ZTF is supported by the National Science Foundation under Grant No. AST-2034437 and a collaboration including Caltech, IPAC, the Weizmann Institute of Science, the Oskar Klein Center at Stockholm University, the University of Maryland, Deutsches Elektronen-Synchrotron and Humboldt University, the TANGO Consortium of Taiwan, the University of Wisconsin at Milwaukee, Trinity College Dublin, Lawrence Livermore National Laboratories, IN2P3, France, the University of Warwick, the University of Bochum, and Northwestern University. Operations are conducted by COO, IPAC, and UW. Some of the data presented herein were obtained at the W.M. Keck Observatory, which is operated as a scientific partnership among the California Institute of Technology, the University of California and the National Aeronautics and Space Administration. This work has made use of data from the European Space Agency (ESA) mission {\it Gaia} \\(https://www.cosmos.esa.int/gaia), processed by the {\it Gaia} Data Processing and Analysis Consortium (DPAC, https://www.cosmos.esa.int/web/gaia/dpac/consortium). Funding for the DPAC has been provided by national institutions, in particular the institutions participating in the {\it Gaia} Multilateral Agreement. The Pan-STARRS1 Surveys (PS1) and the PS1 public science archive have been made possible through contributions by the Institute for Astronomy, the University of Hawaii, the Pan-STARRS Project Office, the Max-Planck Society and its participating institutes, the Max Planck Institute for Astronomy, Heidelberg and the Max Planck Institute for Extraterrestrial Physics, Garching, The Johns Hopkins University, Durham University, the University of Edinburgh, the Queen's University Belfast, the Harvard-Smithsonian Center for Astrophysics, the Las Cumbres Observatory Global Telescope Network Incorporated, the National Central University of Taiwan, the Space Telescope Science Institute, the National Aeronautics and Space Administration under Grant No. NNX08AR22G issued through the Planetary Science Division of the NASA Science Mission Directorate, the National Science Foundation Grant No. AST-1238877, the University of Maryland, Eotvos Lorand University (ELTE), the Los Alamos National Laboratory, and the Gordon and Betty Moore Foundation.
    The design and construction of HiPERCAM was funded by the European Research Council under the European Union's Seventh Framework Programme (FP/2007-2013) under ERC-2013-ADG Grant Agreement no. 340040 (HiPERCAM). VSD and HiPERCAM operations are funded by the Science and Technology Facilities Council (grant ST/V000853/1). This research has made use of NASA’s Astrophysics Data System and of astropy 
 
 \item[Author Contributions] I.C. reduced the UV and optical data, conducted the spectral and photometric analysis, and is the primary author of the manuscript. K.B.B. performed the period search on ZTF data. I.C., K.B.B., P.M., A.C.R., J.v.R. and Z.P.V. performed the observations with \emph{LRIS} and \emph{CHIMERA}. I.C., K.B.B., P.-E.T., L.F., J.F., B.T.G., J.J.H., J.H., A.K, S.R.K., T.R.M., T.A.P., H.B.R., A.C.R., J.v.R., Z.P.V, S.V., and D.W. contributed to the physical interpretation of the object. P.-E.T. contributed the synthetic spectral models and conducted the analysis to estimate the minimum field strength needed to suppress convection in the white dwarf's atmosphere. J.F. constructed MESA models for the object. P.M. developed a reduction pipeline for the ZTF data and contributed to the analysis. D.P. developed a reduction pipeline for the LRIS data. T.R.M., V.S.D., S.P.L., E.B., A.J.B., M.J.D., M.J.G., P.K., S.G.P., I.P. and D.I.S. were responsible for the operation of HiPERCAM.  R.D., A.D., R.R.L., R.L.R., and B.R. contributed to the implementation of ZTF. M.J.G. is the project scientist, E.C.B. is the survey scientist, T.A.P. is the co-PI and S.R.K. is the PI of ZTF.
 
 \item[Competing Interests] The authors declare that they have no
competing financial interests.

 \item[Correspondence] Correspondence and requests for materials
should be addressed to I.C.~(email: ilariac@caltech.edu).
\end{addendum}

\end{document}